\documentclass[twocolumn]{aastex61}
\usepackage{url}

\newcommand{\Msol}{\mbox{$M_{\sun}$}}

\newcommand{\muas}{\mbox{$\mu$as}}

\newcommand{\muJb}{\mbox{$\mu$Jy~beam$^{-1}$}}

\newcommand{\cmsixpc}{\mbox{cm$^{-6}$~pc}}
\newcommand{\cmthreepc}{\mbox{cm$^{-3}$~pc}}
\newcommand{\Mion}{\mbox{$M_{\rm ion}$}}

\newcommand{\ttunity}{\mbox{$t_{\tau = 1}$}}

\defcitealias{SN86J-1}{SN86J-I} 
\defcitealias{SN86J-2}{SN86J-II}
\defcitealias{SN86J-3}{SN86J-III}
\defcitealias{SN86J-4}{SN86J-IV}

\begin{document}

\title{On the Possibility of Fast Radio Bursts from Inside Supernovae:
  The Case of SN 1986J}

\author{Michael F. Bietenholz}\affiliation{Hartebeesthoek Radio
  Observatory, PO Box 443, Krugersdorp 1740, South
  Africa}\affiliation{Department of Physics and Astronomy, York
  University, Toronto, M3J~1P3, Ontario, Canada}

\author{Norbert Bartel}\affiliation{Department of Physics and
  Astronomy, York University, Toronto, M3J~1P3, Ontario, Canada}

\shortauthors{Bietenholz \& Bartel}

\accepted{to the {\em Astrophysical Journal}}

\begin{abstract}
We discuss the possibility of obtaining Fast Radio Bursts (FRBs) from
the interior of supernovae, in particular SN~1986J\@.
Young neutron stars are involved in many of the
possible scenarios for the origin of FRBs, and it has been suggested
that the high dispersion measures observed in FRBs might be produced
by the ionized material in the ejecta of associated
supernovae.  Using VLA and VLBI measurements of the Type IIn SN~1986J,
which has a central compact component not so far seen in other
supernovae, we can directly observe for the first time radio signals
which originate in the interior of a young ($\sim$30~yr old)
supernova.  We show that at age 30~yr, any FRB signal at $\sim$1~GHz
would still be largely absorbed by the ejecta.  By the time the ejecta
have expanded so that a 1-GHz signal would be visible, the internal
dispersion measure due to the SN ejecta would be below the values
typically seen for FRBs.  The high dispersion measures seen for the
FRBs detected so far could of course be due to propagation through the
intergalactic medium provided that the FRBs are at distances much
larger than that of SN~1986J, which is 10~Mpc.  We conclude that if
FRBs originate in Type II SNe/SNRs, they would likely not become
visible till $60 \sim 200$~yr after the SN explosion.
\end{abstract}


\section{Introduction}
\label{sintro}

Fast Radio Bursts (FRBs) are bursts of radio emission, 0.1 to 10 Jy at
$\sim$1~GHz, which occur on timescales of milliseconds or less.  They
are characterized by high dispersion measures (DM) in the range of 375
to 1700 \cmthreepc.  Their origin is still mysterious \citep[for
  recent reviews, see e.g.,][]{Katz2016b, Petroff+2016,
  Thornton+2013}.
They have mostly been detected with single-dish telescopes with low
(arc-minute) resolution\footnote{Recently, \citet{Caleb+2017} reported
  the interferometric detection of several FRBs using the Molongolo
  Observatory Synthesis Telescope, but the bursts are only localized
  to an area of $\sim$11 arcmin$^2$, too large to reliably identify a
  host galaxy.}, hence counterparts are not readily identified.
Although \citet{Keane+2016} suggested that FRB~150418 was associated
with a radio ``afterglow'' in an elliptical galaxy at redshift, $z
\simeq 0.5$, \citet{WilliamsB2016} and \citet{Vedantham+2016} showed
that the ``afterglow'' emission was probably due to AGN variability
and not to the FRB, and that the exact location of FRB~150418 is thus
still not known.

So far, only one, FRB~121102, has been found to repeat
\citep{Spitler+2016}, and consequently it was possible to localize it,
to dwarf galaxy at $z \simeq 0.193$ \citep{Chatterjee+2017,
  Tendulkar+2017}.  A persistent radio counterpart was identified, and
studied using VLBI by \citet{Marcote+2017}\@.  To date, FRB 121102
remains the only FRB which repeats and the only one that is well
localized.  Its DM, after subtraction of the Galactic component, is
$371 \pm 0.8$~\cmthreepc\ \citep{Scholz+2016}, and it has stayed
constant over a period of $\sim$3~yr.

At the time of writing, 23 FRBs are known\footnote{See
  \url{http://www.astronomy.swin.edu.au/pulsar/frbcat/}
  \citep{Petroff+2016}.}.  Of these, 17 are at Galactic latitudes,
$|b| > 5\arcdeg$.  The DMs of the FRBs at $|b|>5\arcdeg$ are far
higher than expected from our own Galaxy's interstellar medium
\citep[see, e.g.,][]{Thornton+2013, Katz2016b}.  The DMs are therefore
analogously higher than can readily be explained by the host galaxies,
except in the few cases where the host galaxies would be edge on, or
if the FRBs occur only very near the galactic centers.  The high DMs
therefore suggest cosmological distances, with very long path-lengths
through the intergalactic medium (IGM), large DMs intrinsic to the
sources within the host galaxies, or a combination of both.

Although a number of arguments have been advanced that FRBs are indeed
at cosmological distances, and that the large DMs are in fact due to
the long path-length through the intergalactic medium, there remain
some difficulties with this scenario, in particular in reproducing the
temporal scattering \citep[see discussion in][]{ConnorSP2016}.

One way of obtaining large intrinsic DMs without recourse to
cosmological distances was proposed by \citet{ConnorSP2016}.  It is
that we are seeing FRBs through the dense ionized ejecta in young
core-collapse supernovae (SNe), with the ionized parts of the dense SN
ejecta and possibly the circumstellar environment providing the
observed dispersion.  The FRBs themselves would be generated by the
pulsars or magnetars left behind by the supernova explosion in the
centers of the clouds of expanding SN ejecta.

We have observed such a nearby core-collapse supernova, SN~1986J, in
the radio, and have obtained both very long baseline interferometry
(VLBI) imaging, and broadband measurements of its spectral energy
distribution (SED) using the National Radio Astronomy
Observatory's\footnote{The National Radio Astronomy Observatory, NRAO,
  is a facility of the National Science Foundation operated under
  cooperative agreement by Associated Universities, Inc.} Karl
G. Jansky Very Large Array (VLA)\@.  SN~1986J is unique among SNe in
having a central component, a bright, compact radio source most likely
located in the very center of the expanding shell of debris.  Although
no FRBs have been associated with SN~1986J, it is important in the
context of FRBs because SN~1986J represents the first case where we
have direct observational constraints on the propagation of
GHz-frequency radio signals through the ejecta of a supernova.  In
this paper, we use our observations of SN~1986J to address the
possibility of propagating an FRB signal through the ejecta of a
30-year old supernova.

\section{SN 1986J and its Central component: Radio Emission from Inside a Young Supernova}
\label{scentral}

Although many supernovae (SNe) are detected optically every year,
optical observations cannot resolve the expanding clouds of ejecta
beyond the Magellanic clouds.  Only VLBI observations have the
necessary resolution, and only a handful of SNe are sufficiently radio
bright and nearby so that resolved images can be obtained.  SN~1986J
was one of the first SNe observed with VLBI \citep{Bartel+1987,
  Bartel+1991}.  It was also one of the most radio-luminous ever
observed, and one of the few SNe still detectable more than $t = 30$
years after the explosion, thus we have been able to follow its
evolution for longer than was possible for most other SNe.  We
describe the VLBI observations of SN~1986J along with VLA observations
to monitor the evolution of its radio SED in a series of papers:
\citet{SN86J-1, SN86J-2}, and \citet{SN86J-3, SN86J-4}, which we will
refer to as SN86J-I to SN86J-IV respectively.  We refer the interested
reader particularly to a sequence of VLBI images in
\citetalias{SN86J-3} which show both the expansion and the
non-selfsimilar evolution over almost three decades and to a
discussion of the evolution of the broadband radio SED in
\citetalias{SN86J-4}.

SN~1986J occurred in the nearby galaxy NGC~891, for whose distance the
NASA/IPAC Extragalactic Database (NED) lists 19 measurements with a
mean of $10.0 \pm 1.4$~Mpc, which value we adopt throughout this
paper.  It was classified as being a Type~IIn supernova
\citep{Rupen+1987}, showing signs of strong interaction of the
expanding supernova shock with the circumstellar material (CSM).

The structure seen in the cm-wavelength VLBI images shows an
expanding, albeit somewhat distorted shell, but also two strong
compact enhancements of the brightness: one, now fading, to the NE of
the shell center, and a second, which remains bright, at or very near
the projected center.  Such a cm-wavelength central radio component
has not so far been seen in any other supernova\footnote{We note that
  central emission at mm wavelengths, but not at cm wavelengths, has
  been seen in SN~1987A \citep{Zanardo+2014}. SN~1987A's mm-wavelength
  central component, which is attributed to dust, does not provide any
  useful constraints on the propagation of cm-wave FRB-like signals
  through the ejecta.}  \citep[see e.g.,][]{SNVLBI_Cagliari,
  BartelB2014IAUS}.

\begin{figure}
\centering 
\includegraphics[width=\linewidth]{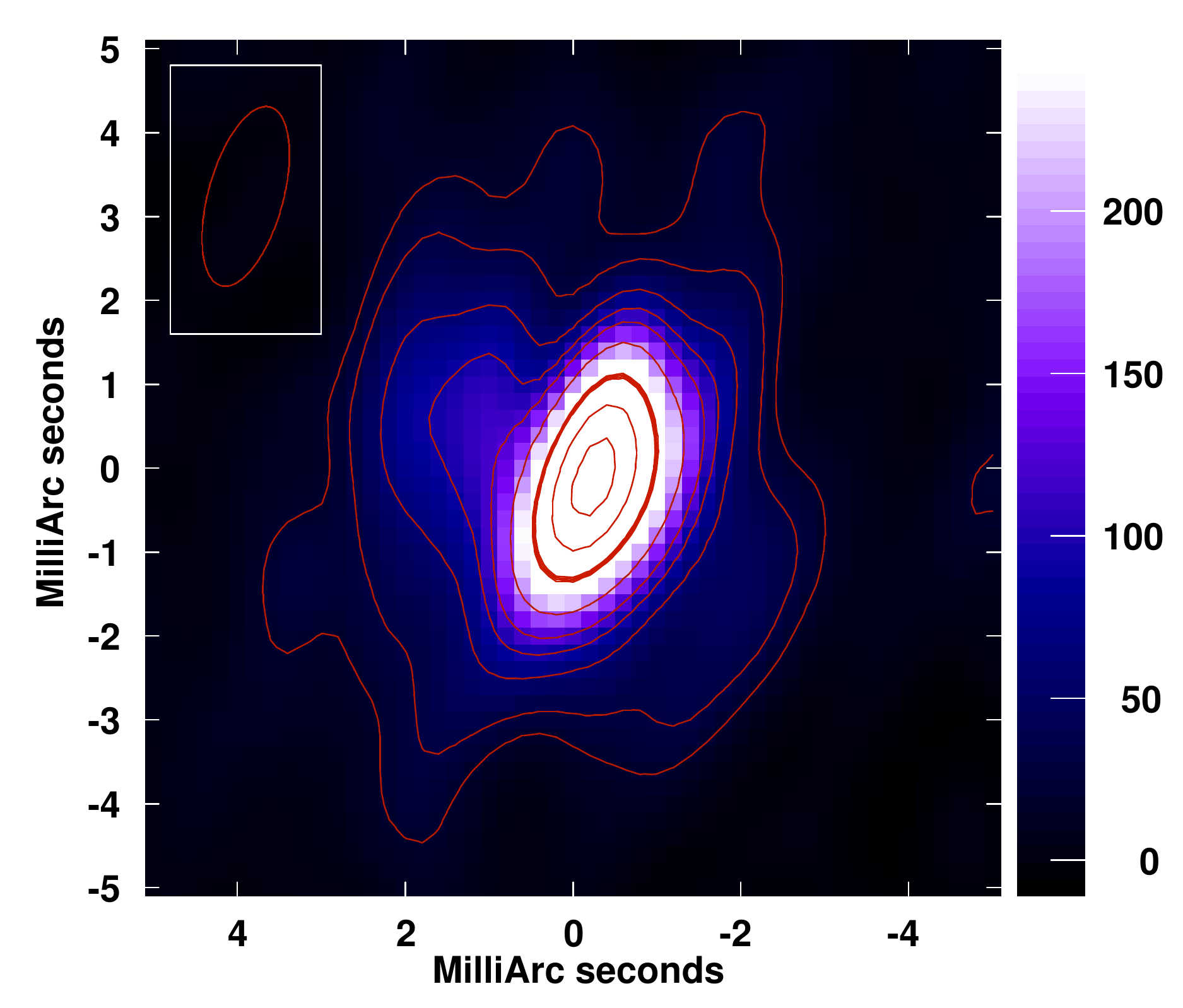}
\caption{The 5-GHz VLBI image of SN~1986J in 2014 ($t = 31.6$~yr),
  reproduced from \citetalias{SN86J-3}.  The contours are drawn at
  $-3$, 3, 5, 10, 15, 20, 30, {\bf 50}, 70 and 90\% of the peak
  brightness, with the 50\% contour being emphasized.  The peak
  brightness was 617~\muJb\ and the background rms brightness was
  5.9~\muJb.  The color-scale is labeled in \muJb.  North is up and
  east to the left, and the FWHM of the convolving beam is indicated
  at upper left.  The image is dominated by the marginally-resolved
  central component.  We argue in \citetalias{SN86J-3, SN86J-4} that
  the central component must be radio emission originating in the
  three-dimensional center of SN~1986J, and not fortuitously central
  only in projection.}
\label{fsn86jimg}
\end{figure}

We showed by means of phase-referenced multi-frequency VLBI imaging,
that the central component in the images was associated with an
inversion which appeared in the SED of SN~1986J in
\citet{SN86J-Sci}\@.  We first saw the central component in 2002 at
15~GHz, and it was not yet visible at 5-GHz. Since then, however, it
has increased steadily in brightness at 5-GHz, and now it dominates
the 5-GHz image.  We show the most recent 5-GHz VLBI image of SN~1986J
at $t = 31.6$~yr, reproduced from \citetalias{SN86J-3}, in
Figure~\ref{fsn86jimg}.

\begin{figure}
\centering 
\includegraphics[width=\linewidth]{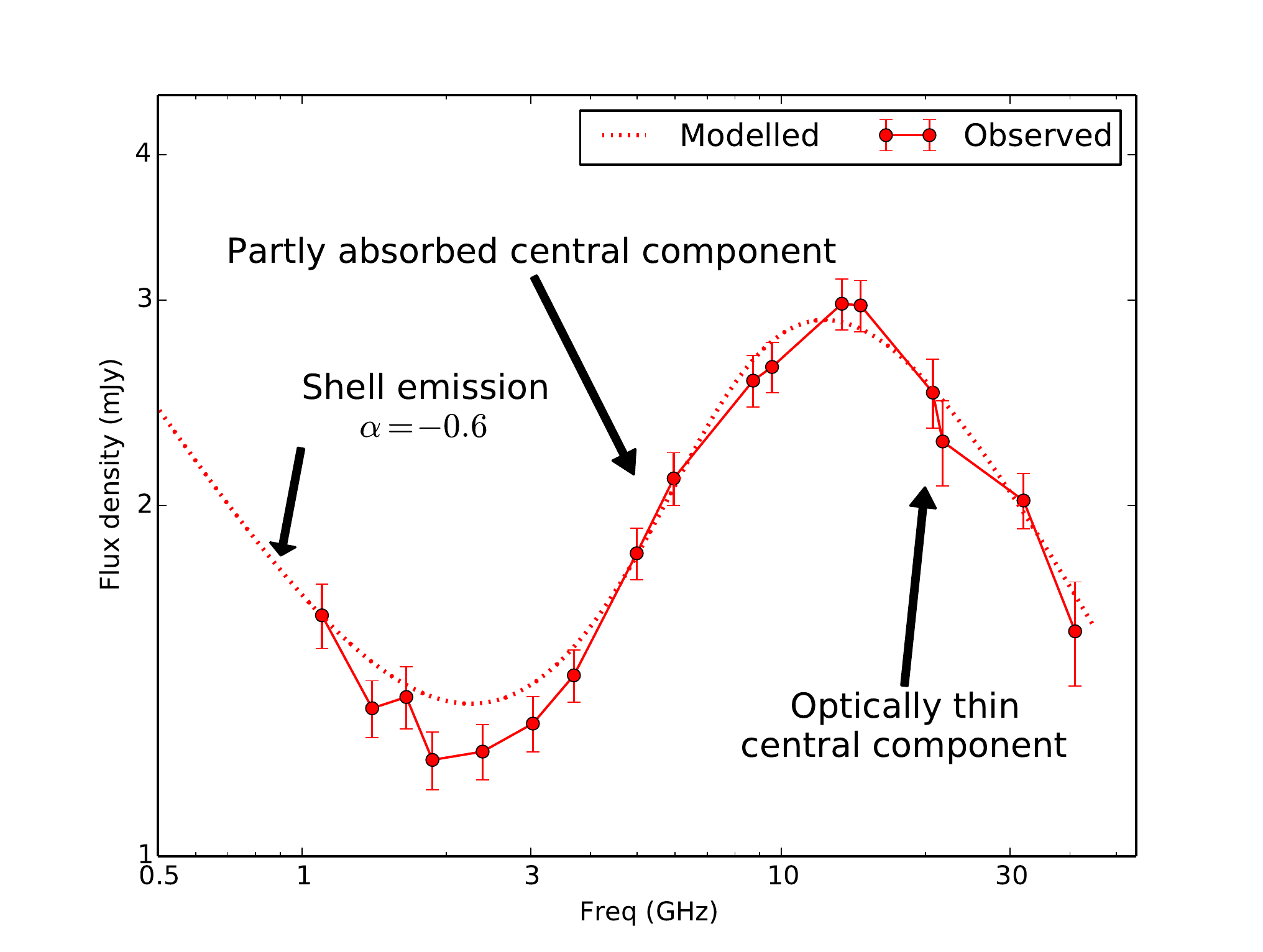}
\caption{The radio spectral energy distribution of SN~1986J on 2012
  April 10 (at age $t = 29.6$~yr).  The points show the VLA flux
  density measurements and their standard errors, and the dotted line
  shows the fitted model of the evolving SED at this epoch.  With time
  the absorption decreases and both the inflection points in the
  spectrum move to lower frequencies.  Details of the model are given
  in \citetalias{SN86J-4}.  The parts of the spectrum are indicated.
  The parameter, $\alpha$, is the spectral index.}
\label{fsed}
\end{figure}

We show the radio SED of SN~1986J at about the same age ($t =
29.6$~yr) in Figure~\ref{fsed} \citepalias[for details, and the
  detailed evolution of the SED, see][]{SN86J-4}.  The part of the SED
due to the central component, i.e., that above $\sim$3~GHz, suggests
that the emission is absorbed below a frequency of $\sim$15~GHz.
Although both synchrotron self-absorption and free-free absorption are
seen commonly in SNe, we argued in \citetalias{SN86J-2} and
\citet{SN86J-COSPAR} that synchrotron-self-absorption is not plausible
in this case.  At $t = 32$~yr, the angular size of the central
component is $900^{+100}_{-500}$~\muas, corresponding at 10 Mpc to a
radius, $r_{\rm comp} = 6.7^{+0.7}_{-3.7} \times 10^{16}$~cm, and the
peak in the SED was 3~mJy at 13~GHz \citepalias{SN86J-4}.  Following
\citet{Chevalier1998}, for that flux density, we can compute that
synchrotron self-absorption would only be important for radii of $<
0.2 \times 10^{15}$~cm.  The central component, therefore, is too
large and faint to suffer significant self-absorption, and we can
therefore conclude that the absorption is dominated by free-free
absorption.

Significant free-free absorption is indeed expected if the central
component is in the center of the supernova and we are seeing it {\em
  through}\/ the shell of expanding ejecta, which is expected to be at
least partly ionized and thus to provide free-free absorption at radio
frequencies.

The longevity, still-increasing flux density at 5-GHz and stationarity
of the central component argue convincingly that it is due to emission
originating in the three-dimensional center of the supernova, rather
than being associated with the expanding shell of ejecta and central
only in projection \citepalias{SN86J-3, SN86J-4}.  The absorption seen
is therefore almost certainly due to the ionized portion of the
intervening SN ejecta.

The central component therefore represents synchrotron radio emission
most likely coming from {\em inside}\/ a young supernova, which has
not so far been observed for any other supernova.  This fact is of
particular interest in the context of FRBs because it gives us direct
observational constraints on the propagation of radio signals through
SN ejecta.

Although SN~1986J was of type IIn, characterized by strong CSM
interaction, the CSM at this late stage is likely to have little
effect on either the absorption or the dispersion.
\citet{WeilerPS1990} show that the 1.4~GHz lightcurve of SN~1986J
reached its peak around $t \sim 5$~yr, therefore the optical depth of
the CSM at 1.4 GHz was $\sim$1 at that time.  Since the supernova has
expanded by at least a factor of 3 since then, the remaining opacity
due to the CSM even at 1~GHz is expected to be quite small.  Since the
densities in the CSM are expected to be low compared to those in the
ejecta, we expect the contribution of the CSM to the DM also to be
small.  Therefore our conclusions should be applicable to different
kinds of Type II SNe, not just to Type IIn's like SN~1986J.

\section{Absorption of Radio Signals by the SN Ejecta}
\label{sabsorb}

As we have shown (see Figure~\ref{fsed}), at $t=29.6$~yr, the emission
from SN~1986J's central component is absorbed below a turnover
frequency of $\sim$15~GHz, almost certainly by free-free absorption.
The optical depth to free-free emission is given by:
$$\tau    =  3.28 \times 10^{-7} \, 
  \left(\frac{\nu}{\mathrm{GHz}}\right)^{-2.1} \, 
  \left(\frac{T_e}{10^4 \, \mathrm{K}}\right)^{-1.35}\, 
  \left(\frac{\mathrm{EM}}{\cmsixpc}\right).$$
where $\nu$ is the frequency, $T_e$ is the electron temperature, and
EM the emission measure.  EM is given by
$$\mathrm{EM} = \int{N_e}^2 dl,$$ where $N_e$ is the number density of
free electrons, and $l$ is the path-length along the line of sight from
the center of the SN to the observer.

The turnover frequency is approximately the frequency where $\tau = 1$
(the exact value depends on the unabsorbed spectral index of the
emission).  So, for that particular time of $t = 29.6$~yr, assuming
that $\tau = 1$ occurs at exactly 15~GHz, we can calculate
that $$\mathrm{EM}\simeq 9\times10^{8} \; \cmsixpc.$$

The SED shown in Figure~\ref{fsed}, however, is not static, since we
found in \citetalias{SN86J-2} and \citetalias{SN86J-4} that the
turnover frequency is decreasing with time.  Time variability
therefore needs to be considered. In order to characterize the
evolving SED of SN~1986J, we fitted a model with 8 free parameters to
the flux density measurements (such as those in Fig.~\ref{fsed})
between $t=14$ and 30~yr using Bayesian statistics.  Our model
consisted of a shell emission component, which is optically thin, as
expected at these late times, and a central component which is partly
absorbed, with both the intrinsic flux densities and the amount of
absorption varying with time in a powerlaw fashion. Note that in our
model, the SED is calculated accurately, and we did not make the
simplifying assumption above that the spectral peak occurs exactly at
the frequency where $\tau = 1$.  Details of our fit are given in
\citetalias{SN86J-4}. The fit adequately reproduces the downward
evolution of the turnover frequency with time as well as the shape of
the SED at any given time.

Of interest here is the result on the absorption.  The relevant fitted
parameter in our model is the emission measure (EM) as a function of
time. We found that
$$ {\rm EM} = (1.64 \pm 0.21) \times 10^9 \, (t/{\rm \, 20 \,
  yr})^{-2.72 \pm 0.26} \; \cmsixpc . $$ 
The EM, therefore is decreasing in time.  This is expected as the SN
is expanding with time.  In a homologously expanding system with a
constant number of free electrons and $r \propto t^{q}$, one would
expect EM to be proportional to $t^{-5q}$, so the fitted
time-dependence of EM suggests a system expanding with $r \propto
t^{0.54\pm0.05}$.  This is somewhat more decelerated than the forward
shock
of SN~1986J, which we found to be expanding with $r \propto
t^{0.69\pm0.03}$ \citepalias{SN86J-2}, suggesting perhaps some ongoing
fragmentation of the ejecta, which could cause the opacity to not
scale simply with the radius.

This result on the EM and its time-dependence gives us a unique
observational constraint on the column density of free electrons
between us and the center of SN~1986J, and thus allows us to put some
constraints on the time that SN ejecta would become transparent to
FRB-like signals (i.e., $\sim$1~GHz), and on the dispersion measure.

\section{Time Until SN Becomes Transparent to an FRB Signal}
\label{sabsFRB}

Given that, at the present time ($t = 29.6$~yr), the turnover
frequency is at $\sim$15~GHz (Fig.~\ref{fsed}), it is clear that any
FRB-like signals, which are generally observed in the 0.6 to 1.4~GHz
range \citep[e.g.,][]{Katz2016b}, originating in the center of
SN~1986J would still be heavily absorbed.

Extrapolating our fit to the evolving value of EM$(t)$ above, we
obtain a best estimate of the age of the SN at which the optical depth
at 1~GHz would reach unity of $\ttunity \, = 200$~yr.
The uncertainty on this best estimate is not easy to compute: the
model we fitted to SN~1986J's SED was only approximate, and it was
fitted to measurements only between $t = 15$ to 30~yr, thus the
extrapolation to times $t \gg 30$~yr is quite uncertain.  Nonetheless,
it is clear that the ejecta would not be transparent to signals at
typical FRB frequencies of $\sim$1~GHz until at least three decades
from now, and we take a lower bound of $\ttunity > 60$~yr.

\section{Dispersion Measure}
\label{sDM}

One of the distinguishing features of FRBs is their high values of the
dispersion measure,
$$ DM = \int{N_e \cdot dl}\,.$$ Can we calculate the DM for SN~1986J
from our estimate of the EM, and could such high values of DM still be
produced locally by the ejecta of by the time the ejecta have become
transparent to 1-GHz radiation?

We found in \S~\ref{sabsFRB} that the expected optical depth at 1~GHz
of SN~1986J would reach unity at $\ttunity \sim$200~yr, at which time
the EM would be $3 \times 10^6$~\cmsixpc, and the extrapolated outer
radius of the SN would be $\sim$0.9~pc \citepalias{SN86J-2}.

To calculate the DM from the EM we need to know the distribution of
the free electrons along the line of sight.  In a supernova, a dual
shock structure will form with a forward shock being driven into the
CSM and a reverse shock being driven back into the ejecta. We show a
schematic diagram of the structure of the SN in
Figure~\ref{fdiagram}. The thickness of the region between the shocks
is $\sim$20\% of the forward-shock radius
\citep[e.g.,][]{Chevalier1982b, SN93J-3}.  The material between the
two shocks is at sufficiently low density and high temperatures
\citep[$T_e > 10^6$~K; e.g.,][]{Chevalier1982b, LundqvistF1988} that
it does not contribute significantly to the radio absorption, or to
our values of EM\@.

Due to the low density of the region between the shocks, this region
is also not expected to contribute to the DM\@.  For example, for an
outer shock radius of 0.90~pc and an inner shock radius of 0.72~pc, a
completely ionized mass of 10~\Msol\ between the two shocks would
produce a DM of only $\sim$40~\cmthreepc.  Since the mass between the
two shocks will almost certainly be less than 10~\Msol, we also ignore
the contribution to the DM from the region between the two shocks.  We
need consider, therefore, only the unshocked ejecta inside the reverse
shock, and of these it is only the ionized fraction which contributes
significantly to the the EM or DM or both.

However, that ionized fraction and its distribution within the ejecta
is not well known.  It is generally expected that the ejecta are
largely ionized by the SN shock breakout, however recombination may be
rapid thereafter.  While some portions of the ejecta may recombine,
the radiation from the shocks will likely keep at least the outer
portion of the ejecta ionized.  We therefore consider various
distributions of ionized material within the ejecta.  In particular,
we consider three example distributions of the ionized material, which
are illustrated schematically in Figure~\ref{fdiagram}, namely: case
``A'', where the ejecta are uniformly ionized, case ``B'', where they
are ionized from the outside, but are neutral towards the center of
the SN, and case ``C'', where they are ionized from the center of the
SN, but neutral towards the outside.

We seek to constrain the DM based on the observed absorption, via our
fitted values of EM\@.  For our calculation, we use the extrapolated
time when the SN becomes optically thin at 1~GHz of \ttunity\, =
200~yr, along with the extrapolated value of EM of
$3\times10^6$~\cmsixpc\ that we get for that time (\S~\ref{sabsFRB}).
We further take, at that time, the radius of the forward shock to be
0.9~pc, and that of the reverse shock to be 80\% of that, or 0.72~pc,
again extrapolating our the measured expansion of SN~1986J to $t =
200$~yr \citepalias{SN86J-2}.

\subsection{Uniform Distribution of the Ionized Ejecta}
\label{suniform}

\begin{figure}
\centering 
\includegraphics[width=\linewidth]{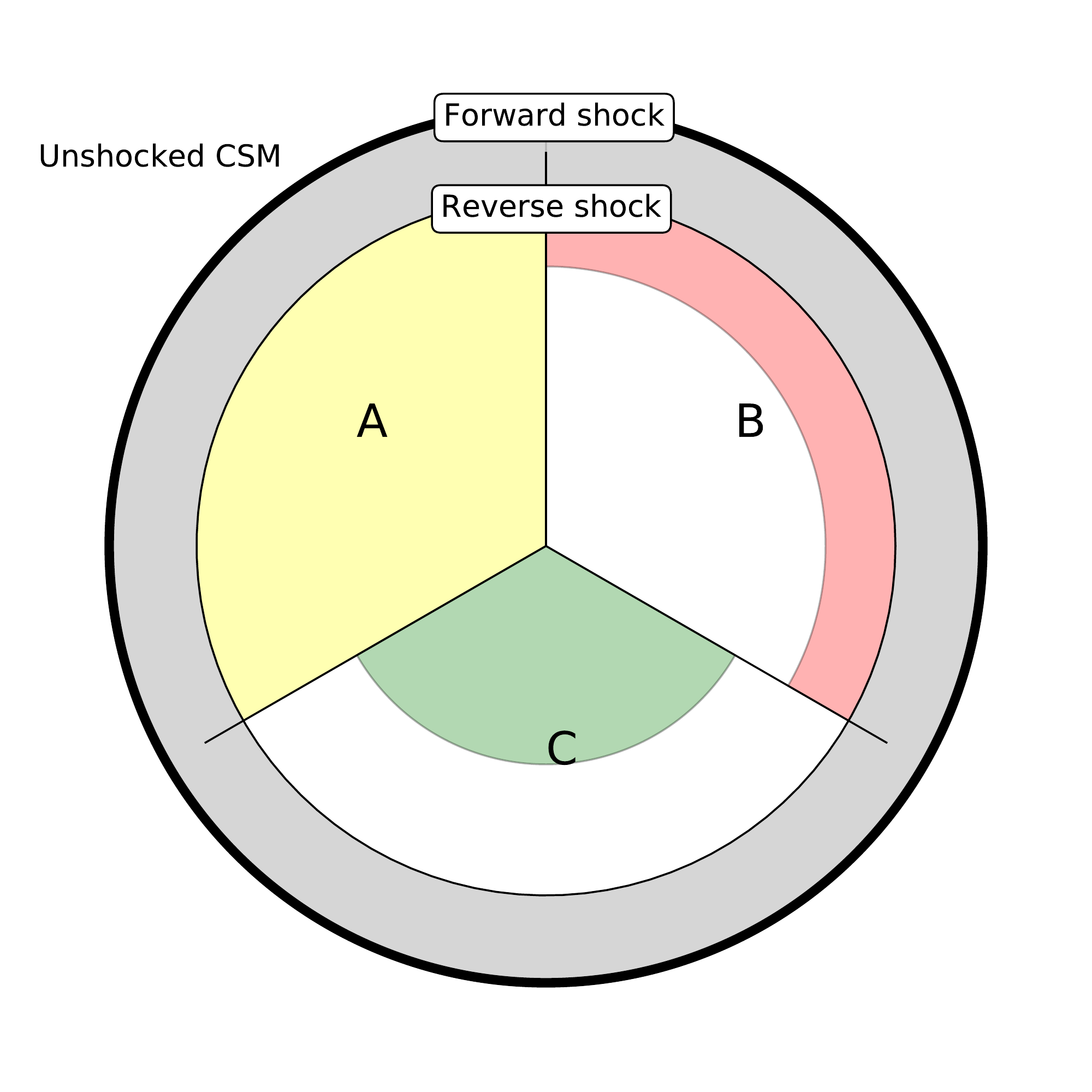}
\caption{A schematic illustration of the different possible
  distributions of ionized material within the SN ejecta.  The forward
  and reverse shocks are indicated, with the former being the heavy
  outer circle.  The three sectors labeled A through C show the three
  different distributions of ionized material discussed in the text,
  with the ionized material being shaded.  Case A (yellow) represents
  fully ionized ejecta.  Case B (red) represents ejecta which have
  become neutral, but are ionized from the outside by the shocks.
  Case C (green) indicates ejecta ionized near the center of the SN,
  but neutral in the outer regions towards the reverse shock.  The
  material between the two shocks, shown in gray, is expected to be
  fully ionized but to contribute little to either the absorption (due
  to the high temperature) or the dispersion (due to low density) is
  shown in gray.  The unshocked CSM is exterior to the forward shock.}
\label{fdiagram}
\end{figure}

For the first and simplest case, we assume that the ionized ejecta are
uniformly distributed inside the reverse shock, which is sketched as
case A in Figure~\ref{fdiagram}.  We therefore have a sphere of
uniform $N_e$ out to $r = 0.7$~pc.  From our extrapolated value of EM,
we can straightforwardly calculate that, at $t = 200$~yr, $N_e =
2.0\times10^3$~cm$^{-3}$ and DM $=1.3\times10^3$~\cmthreepc.

While this value of DM is compatible with the values of 375 to
1700~\cmthreepc\ seen for FRBs, if we assume fully ionized material
with a mass in u (unified atomic mass units) per free electron of
$\mu_e = 1.3$, it implies an unreasonable total ionized mass, \Mion =
100 \Msol.  Therefore, to produce the EM with only reasonable masses,
the ionized material would have to be non-uniformly distributed.

\subsection{Non-Uniform Distribution of the Ionized Ejecta}
\label{snonuniform}

The value of DM corresponding to a particular one of EM depends only
on the path-length through the ionized portion of the material (equal
to the reverse-shock radius in the case of a sphere of uniform $N_e$
assumed for case A in \S~\ref{suniform} above).  We can therefore
consider cases B, ejecta ionized from the outside, and C, ejecta
ionized from the inside, together for the calculation of the DM\@.  We
compute the values of DM corresponding to different path-lengths, for
$\ttunity\, = 200$~yr, and the same outer shock radius and
extrapolated value of EM used in \S~\ref{suniform} above (0.9~pc and
$3 \times 10^6$~\cmsixpc\ respectively).  In case B, the path-length
is the thickness of the ionized shell, while in case C it is the
radius of the ionized sphere.

\begin{figure}
\centering \includegraphics[width=\linewidth]{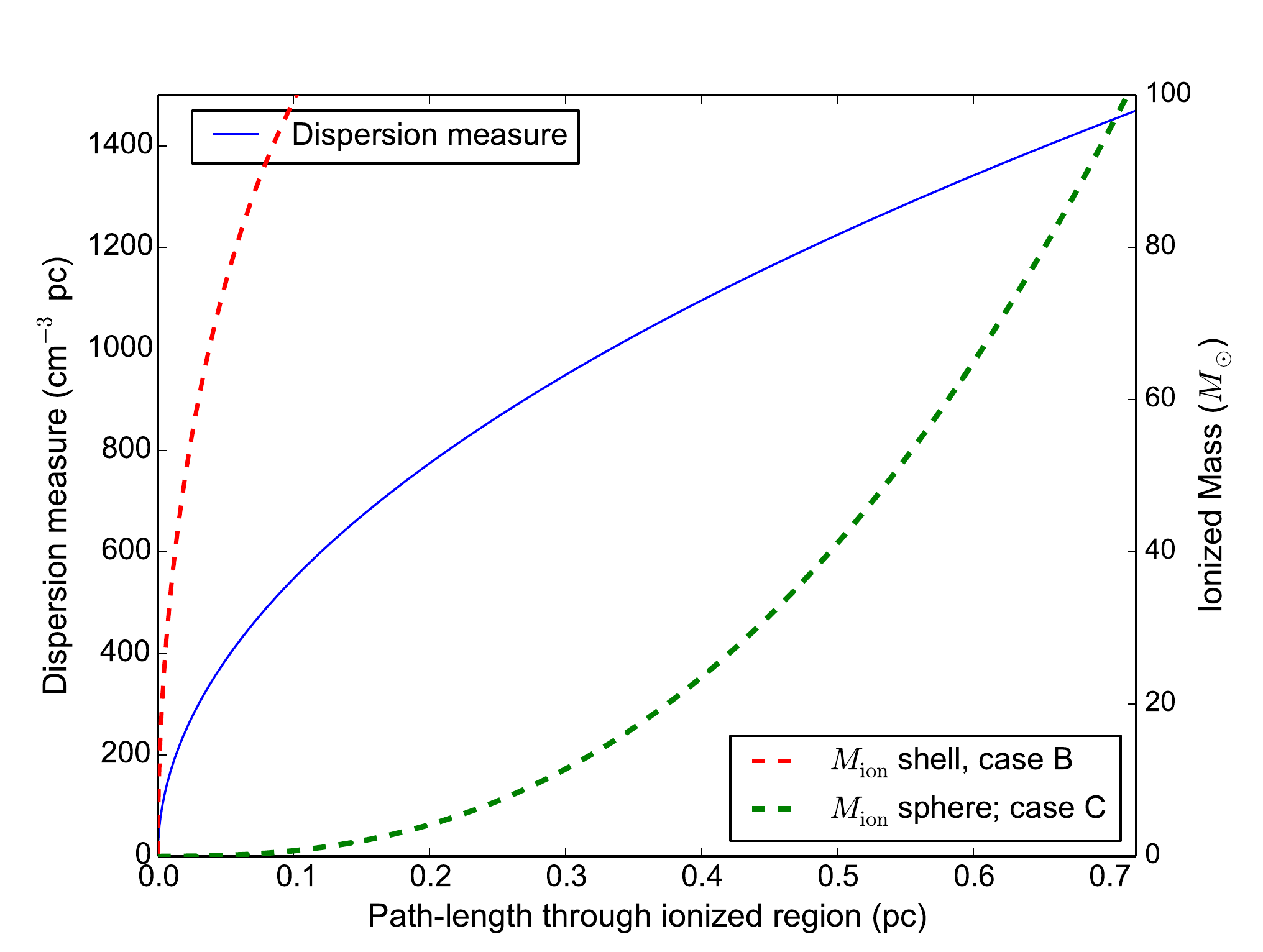}
\caption{The dispersion measure (DM) and ionized mass (\Mion) through
  the ejecta of SN~1986J for an emission measure (EM) of $3 \times
  10^6$~\cmsixpc, the value at which the optical depth for 1~GHz is
  unity.  For SN~1986J, this value of the EM is reached at $\ttunity
  \sim 200$~yr.  The blue solid line and the left vertical axis show
  the dispersion measure (DM) as a function of the path-length through
  the ionized region.  The right vertical axis and the two dotted
  lines show the total ionized mass, \Mion, calculated assuming 1.3 u
  per electron.  We show \Mion\ for two different distributions of
  ionized material.  The first is a spherical shell extending inwards
  from an outer radius 0.72~pc (case B), where the path-length is the
  thickness of the shell.  The second is a sphere, where the
  path-length is equal to the radius (case C).  In either case,
  unreasonably large values of $\Mion$ are required to produce
  FRB-like values of DM (300 to 1700 \cmthreepc).}
\vspace{0.1in}  
\label{fDM}
\end{figure}

We plot the values of DM in Figure~\ref{fDM}.  Values of DM in the
range observed for FRBs can indeed be achieved for path-lengths of
0.05 to 1 pc.  The question however remains whether these high values
of DM are consistent with the expected values of \Mion\ in a SN.

As in \S~\ref{suniform}, we calculate the total ionized mass, \Mion,
associated with cases B and C, again assuming $\mu_e = 1.3$\@.  Unlike
DM, \Mion\ does depend on the distribution of ionized matter and not
just on the path-length, and it therefore differs between cases B and
C, so we plot one curve for each of the two cases in Figure~\ref{fDM}.
We note that this figure is very similar to Figure~5 of
\citetalias{SN86J-4}, but that figure was for $t = 20$~yr, whereas in
the present one we plot the values for the extrapolated time of
transparency of $t \sim 200$~yr.

In both cases, for large values of DM, unrealistically large values of
\Mion\ are required.  What is a reasonable upper limit for \Mion?
Although Type II SNe could easily have more than 10~\Msol\ of total
ejected material, only a fraction is expected to be ionized at late
times (decades or more after the explosion).  The expected values for
SN~1986J are likely in the range of 0.5 to 5~\Msol.  In the case of
SN~19867A, \citet{Zanardo+2014} argue that \Mion\ in the ejecta is in
the range of 0.7 to 2.5~\Msol.  Since SN~1986J's progenitor was
probably more massive than that of SN~1987A, we consider values of
\Mion\ up to 5~\Msol.  The generally expected picture is that the
ejecta will cool and become neutral except for a shell which is heated
by emission from the shocks \citep{HamiltonS1984, ChevalierF1994},
which corresponds to our case B, and in which case \Mion\ is expected
to be $<25$\% of the total ejecta mass.

For a spherical shell distribution with outer radius, $r_{\rm outer}$,
equal to the reverse shock radius (0.72~pc, case B), which is what is
expected if the ejecta are ionized from the outside by the shocks, a
mass of 5~\Msol\ is reached already at DM of only 25~\cmthreepc, with
the shell thickness being quite thin at $2\times10^{-4}$~pc or 0.03\%
of the forward shock radius.  Such a thin shell containing a
substantial fraction of the total ejecta mass seems implausible.  We
note that even if \Mion\ is as high as 10~\Msol, the DM reaches only
50~\cmthreepc\ with the shell thickness still only being
$9\times10^{-4}$ pc.

In fact we had already found in \citetalias{SN86J-4} that the required
EM can only be produced if \Mion\ is concentrated near the center of
the SN, rather than being out near the reverse shock (or of course if
the distribution is strongly aspherical).  If \Mion\ is indeed
distributed in a spherical region near the center of SN~1986J (case
C), then a DM of 800~\cmthreepc\ is produced when \Mion\ =
5~\Msol\ and the radius of the sphere is 0.2~pc.  Such a distribution
may in fact be possible for SN~1986J, where the central component may
ionize the ejecta from the center, but is unlikely for other SNe which
do not have any central component.

We can conclude therefore, that by the time the supernova ejecta have
become transparent to radio waves at 1~GHz, it is difficult to produce
the high values of DM typical of FRBs from the supernova ejecta
without either an implausibly high \Mion\ or having the ionized ejecta
concentrated in an implausibly small region.

We note that a large value of DM with a reasonable \Mion\ in a shell
distribution can be produced with a thin shell of small radius. This
case is similar to the spherical distribution with a small radius. One
possibility for producing such a shell inside an expanding SN would be
if a pulsar wind nebula (PWN), with a radius much smaller than that of
the SN forward shock, re-ionizes the ejecta from the inside. In that
case a thin, dense shell could produce a value of DM of
1000~\cmthreepc\ (for our extrapolated EM of $3\times10^6$~\cmsixpc)
with only a reasonable mass.  Although SN~1986J's central component
could be interpreted as being a young PWN nebula, we argue in
\citetalias{SN86J-4} that a PWN hypothesis is not favored.

Regardless of the exact distribution and location of the ionized
portion of the ejecta, if DM were produced by propagation through
the expanding SN ejecta, one would expect it to decrease relatively
rapidly as the ejecta expanded, since DM is proportional to $r^{-2}$
in an expanding but otherwise unchanging system
\citep[e.g.,][]{Piro2016, KashiyamaM2017, Yang+2017}
SNe typically expand with $r \propto t^{0.6 \sim 1}$ so one would
expect that DM $\propto t^{-(1.2 \sim 2)}$.  The repeating FRB~121102
has been observed for over 3~yr, and with the DM changing by $<
10$~\cmthreepc\ or $<3$\% over that period \citep{Scholz+2016,
  YangZ2017, Law+2017}.
However, given its distance ($z = 0.193$) the proportion of the DM due
to the IGM probably dominates, and only a relatively small fraction
would be intrinsic and time-variable due to expanding SN
ejecta.\footnote{\citet{Yang+2017} derive a value of the host-galaxy
  DM of FRB~121102 of $\sim$210 \cmthreepc, but that value is based on
  calculating a value of the intergalactic DM for the known redshift
  of $z = 0.193$ of 164~\cmthreepc\ based on the work of
  \citet{McQuinn2014}, but the latter also shows that the dispersion
  in the value of the intergalactic DM over different lines of sight
  is expected to be $100 \sim 200$~\cmthreepc, and the uncertainty in
  the host-galaxy value of DM for FRB~121102 must therefore be at
  least as large.}
Our measurements of SN~1986J show that ejecta do not become
transparent to FRB-like emission ($\sim 1$~GHz) till $t \simeq 60 -
200$~yr after the explosion, and thus the fractional time-variability
of the intrinsic DM over a few years would not be large.

\section{Summary and Conclusions}

Our VLA and VLBI observations of the core-collapse Type IIn SN~1986J,
which has a bright, central radio component, have for the first time
given us direct observational constraints on the propagation of radio
signals through the ejecta of a young SN\@.  Based on our results from
SN~1986J, we conclude that FRB signals at $\sim$1~GHz would not be
able to propagate through SN~1986J's ejecta for some decades, and that
by the time they could, it would be difficult for the SN ejecta to
produce the high values of DM seen in FRBs. This echoes the
conclusions of \citet{Piro2016}, \citet{MuraseKM2016},
\citet{PiroB2017}, and \citet{MetzgerBM2017}, who all concluded that
Type II SN ejecta would be opaque to FRB-like signals for periods of
several decades to a century or more except possibly in the case of a
stripped-envelope (Type I b/c) SN\@.  Our conclusion is also in accord
with those of \citet{Katz2016a}, who examined the distribution of DM
from the observed FRBs and concluded that it was inconsistent with
that expected if the DMs were produced by the propagation of signals
through SN ejecta alone.

Our conclusions are not in conflict with the hypothesis that FRBs
originate from young pulsars or magnetars, which are among the more
promising hypotheses for the origins of FRBs \citep[see,
  e.g.,][]{Katz2016b}.  However, we find on the basis of observational
evidence that the FRB signals would likely be absorbed for the first
few decades for a supernova similar to SN~1986J, i.e.\ of Type II, and
that the high DMs are unlikely to be caused by propagation of the FRB
signal through the SN ejecta. Very likely therefore, some other
explanation for the large DMs, such as cosmological distances, must be
sought.

\vspace{10pt}
\noindent In summary:
\begin{trivlist}

\item{1.} Our observations of radio emission from SN~1986J's central
  component show that FRB-like signals from inside a Type II SN (at
  $\nu \sim$1~GHz) would be free-free absorbed by the ionized material
  in the ejecta for periods of $60 \sim 200$~yr after the supernova
  explosion.

\item{2.} Once the ejecta have become optically thin to 1-GHz
  radiation, producing the dispersion measures required for FRBs (375
  to 1700 \cmthreepc) requires either an unrealistically large mass of
  ionized material in the ejecta ($> 5 \Msol$) or the confinement of
  the ionized portion of the ejecta to an implausibly small region
  within the SN.

\end{trivlist}

\section*{Acknowledgments }

We have made use of NASA's Astrophysics Data System Abstract Service,
as well as the NASA/IPAC Extragalactic Database (NED) which is
operated by the Jet Propulsion Laboratory, California Institute of
Technology, under contract with the National Aeronautics and Space
Administration. This research was supported by both the National
Sciences and Engineering Research Council of Canada and the National
Research Foundation of South Africa.

\bibliographystyle{aasjournal} 
\bibliography{mybib1,FRB,sn86j_IV_temp}

\begin{thebibliography}{}
\expandafter\ifx\csname natexlab\endcsname\relax\def\natexlab#1{#1}\fi

\bibitem[{{Bartel} \& {Bietenholz}(2014)}]{BartelB2014IAUS}
{Bartel}, N., \& {Bietenholz}, M.~F. 2014, in IAU Symposium, Vol. 296,
  Supernova Environmental Impacts, ed. A.~{Ray} \& R.~A. {McCray}, 53--57

\bibitem[{{Bartel} {et~al.}(1991){Bartel}, {Rupen}, {Shapiro}, {Preston}, \&
  {Rius}}]{Bartel+1991}
{Bartel}, N., {Rupen}, M.~P., {Shapiro}, I.~I., {Preston}, R.~A., \& {Rius}, A.
  1991, \nat, 350, 212

\bibitem[{{Bartel} {et~al.}(1987){Bartel}, {Ratner}, {Rogers}, {Shapiro},
  {Bonometti}, {Cohen}, {Gorenstein}, {Marcaide}, \& {Preston}}]{Bartel+1987}
{Bartel}, N., {Ratner}, M.~I., {Rogers}, A.~E.~E., {et~al.} 1987, \apj, 323,
  505

\bibitem[{{Bietenholz}(2014)}]{SNVLBI_Cagliari}
{Bietenholz}, M. 2014, in {12th European VLBI Network Symposium and Users
  Meeting (2014), published by SISSA, Trieste}, ed. A.~{Tarchi},
  M.~{Giroletti}, \& L.~{Feretti}, 51

\bibitem[{{Bietenholz} \& {Bartel}(2017{\natexlab{a}})}]{SN86J-3}
{Bietenholz}, M.~F., \& {Bartel}, N. 2017{\natexlab{a}}, \apj, 839, 10

\bibitem[{{Bietenholz} \& {Bartel}(2017{\natexlab{b}})}]{SN86J-4}
---. 2017{\natexlab{b}}, ArXiv e-prints, arXiv:1707.06596

\bibitem[{{Bietenholz} {et~al.}(2002){Bietenholz}, {Bartel}, \&
  {Rupen}}]{SN86J-1}
{Bietenholz}, M.~F., {Bartel}, N., \& {Rupen}, M.~P. 2002, \apj, 581, 1132

\bibitem[{{Bietenholz} {et~al.}(2003){Bietenholz}, {Bartel}, \&
  {Rupen}}]{SN93J-3}
---. 2003, \apj, 597, 374

\bibitem[{{Bietenholz} {et~al.}(2004){Bietenholz}, {Bartel}, \&
  {Rupen}}]{SN86J-Sci}
---. 2004, Science, 304, 1947

\bibitem[{{Bietenholz} {et~al.}(2005){Bietenholz}, {Bartel}, \&
  {Rupen}}]{SN86J-COSPAR}
---. 2005, Advances in Space Research, 35, 1052

\bibitem[{{Bietenholz} {et~al.}(2010){Bietenholz}, {Bartel}, \&
  {Rupen}}]{SN86J-2}
---. 2010, \apj, 712, 1057

\bibitem[{{Caleb} {et~al.}(2017){Caleb}, {Flynn}, {Bailes}, {Barr}, {Bateman},
  {Bhandari}, {Campbell-Wilson}, {Farah}, {Green}, {Hunstead}, {Jameson},
  {Jankowski}, {Keane}, {Parthasarathy}, {Ravi}, {Rosado}, {van Straten}, \&
  {Venkatraman Krishnan}}]{Caleb+2017}
{Caleb}, M., {Flynn}, C., {Bailes}, M., {et~al.} 2017, \mnras, 468, 3746

\bibitem[{{Chatterjee} {et~al.}(2017){Chatterjee}, {Law}, {Wharton},
  {Burke-Spolaor}, {Hessels}, {Bower}, {Cordes}, {Tendulkar}, {Bassa},
  {Demorest}, {Butler}, {Seymour}, {Scholz}, {Abruzzo}, {Bogdanov}, {Kaspi},
  {Keimpema}, {Lazio}, {Marcote}, {McLaughlin}, {Paragi}, {Ransom}, {Rupen},
  {Spitler}, \& {van Langevelde}}]{Chatterjee+2017}
{Chatterjee}, S., {Law}, C.~J., {Wharton}, R.~S., {et~al.} 2017, \nat, 541, 58

\bibitem[{{Chevalier}(1982)}]{Chevalier1982b}
{Chevalier}, R.~A. 1982, \apj, 259, 302

\bibitem[{{Chevalier}(1998)}]{Chevalier1998}
---. 1998, \apj, 499, 810

\bibitem[{{Chevalier} \& {Fransson}(1994)}]{ChevalierF1994}
{Chevalier}, R.~A., \& {Fransson}, C. 1994, \apj, 420, 268

\bibitem[{{Connor} {et~al.}(2016){Connor}, {Sievers}, \& {Pen}}]{ConnorSP2016}
{Connor}, L., {Sievers}, J., \& {Pen}, U.-L. 2016, \mnras, 458, L19

\bibitem[{{Hamilton} \& {Sarazin}(1984)}]{HamiltonS1984}
{Hamilton}, A.~J.~S., \& {Sarazin}, C.~L. 1984, \apj, 287, 282

\bibitem[{{Kashiyama} \& {Murase}(2017)}]{KashiyamaM2017}
{Kashiyama}, K., \& {Murase}, K. 2017, \apjl, 839, L3

\bibitem[{{Katz}(2016{\natexlab{a}})}]{Katz2016b}
{Katz}, J.~I. 2016{\natexlab{a}}, Modern Physics Letters A, 31, 1630013

\bibitem[{{Katz}(2016{\natexlab{b}})}]{Katz2016a}
---. 2016{\natexlab{b}}, \apj, 818, 19

\bibitem[{{Keane} {et~al.}(2016){Keane}, {Johnston}, {Bhandari}, {Barr},
  {Bhat}, {Burgay}, {Caleb}, {Flynn}, {Jameson}, {Kramer}, {Petroff},
  {Possenti}, {van Straten}, {Bailes}, {Burke-Spolaor}, {Eatough}, {Stappers},
  {Totani}, {Honma}, {Furusawa}, {Hattori}, {Morokuma}, {Niino}, {Sugai},
  {Terai}, {Tominaga}, {Yamasaki}, {Yasuda}, {Allen}, {Cooke}, {Jencson},
  {Kasliwal}, {Kaplan}, {Tingay}, {Williams}, {Wayth}, {Chandra}, {Perrodin},
  {Berezina}, {Mickaliger}, \& {Bassa}}]{Keane+2016}
{Keane}, E.~F., {Johnston}, S., {Bhandari}, S., {et~al.} 2016, \nat, 530, 453

\bibitem[{{Law} {et~al.}(2017){Law}, {Abruzzo}, {Bassa}, {Bower},
  {Burke-Spolaor}, {Butler}, {Cantwell}, {Carey}, {Chatterjee}, {Cordes},
  {Demorest}, {Dowell}, {Fender}, {Gourdji}, {Grainge}, {Hessels}, {Hickish},
  {Kaspi}, {Lazio}, {McLaughlin}, {Michilli}, {Mooley}, {Perrott}, {Ransom},
  {Razavi-Ghods}, {Rupen}, {Scaife}, {Scott}, {Scholz}, {Seymour}, {Spitler},
  {Stovall}, {Tendulkar}, {Titterington}, {Wharton}, \& {Williams}}]{Law+2017}
{Law}, C.~J., {Abruzzo}, M.~W., {Bassa}, C.~G., {et~al.} 2017, ArXiv e-prints,
  arXiv:1705.07553

\bibitem[{{Lundqvist} \& {Fransson}(1988)}]{LundqvistF1988}
{Lundqvist}, P., \& {Fransson}, C. 1988, \aap, 192, 221

\bibitem[{{Marcote} {et~al.}(2017){Marcote}, {Paragi}, {Hessels}, {Keimpema},
  {van Langevelde}, {Huang}, {Bassa}, {Bogdanov}, {Bower}, {Burke-Spolaor},
  {Butler}, {Campbell}, {Chatterjee}, {Cordes}, {Demorest}, {Garrett}, {Ghosh},
  {Kaspi}, {Law}, {Lazio}, {McLaughlin}, {Ransom}, {Salter}, {Scholz},
  {Seymour}, {Siemion}, {Spitler}, {Tendulkar}, \& {Wharton}}]{Marcote+2017}
{Marcote}, B., {Paragi}, Z., {Hessels}, J.~W.~T., {et~al.} 2017, \apjl, 834, L8

\bibitem[{{McQuinn}(2014)}]{McQuinn2014}
{McQuinn}, M. 2014, \apjl, 780, L33

\bibitem[{{Metzger} {et~al.}(2017){Metzger}, {Berger}, \&
  {Margalit}}]{MetzgerBM2017}
{Metzger}, B.~D., {Berger}, E., \& {Margalit}, B. 2017, \apj, 841, 14

\bibitem[{{Murase} {et~al.}(2016){Murase}, {Kashiyama}, \&
  {M{\'e}sz{\'a}ros}}]{MuraseKM2016}
{Murase}, K., {Kashiyama}, K., \& {M{\'e}sz{\'a}ros}, P. 2016, \mnras, 461,
  1498

\bibitem[{{Petroff} {et~al.}(2016){Petroff}, {Barr}, {Jameson}, {Keane},
  {Bailes}, {Kramer}, {Morello}, {Tabbara}, \& {van Straten}}]{Petroff+2016}
{Petroff}, E., {Barr}, E.~D., {Jameson}, A., {et~al.} 2016, \pasa, 33, e045

\bibitem[{{Piro}(2016)}]{Piro2016}
{Piro}, A.~L. 2016, \apjl, 824, L32

\bibitem[{{Piro} \& {Burke-Spolaor}(2017)}]{PiroB2017}
{Piro}, A.~L., \& {Burke-Spolaor}, S. 2017, \apjl, 841, L30

\bibitem[{{Rupen} {et~al.}(1987){Rupen}, {van Gorkom}, {Knapp}, {Gunn}, \&
  {Schneider}}]{Rupen+1987}
{Rupen}, M.~P., {van Gorkom}, J.~H., {Knapp}, G.~R., {Gunn}, J.~E., \&
  {Schneider}, D.~P. 1987, \aj, 94, 61

\bibitem[{{Scholz} {et~al.}(2016){Scholz}, {Spitler}, {Hessels}, {Chatterjee},
  {Cordes}, {Kaspi}, {Wharton}, {Bassa}, {Bogdanov}, {Camilo}, {Crawford},
  {Deneva}, {van Leeuwen}, {Lynch}, {Madsen}, {McLaughlin}, {Mickaliger},
  {Parent}, {Patel}, {Ransom}, {Seymour}, {Stairs}, {Stappers}, \&
  {Tendulkar}}]{Scholz+2016}
{Scholz}, P., {Spitler}, L.~G., {Hessels}, J.~W.~T., {et~al.} 2016, \apj, 833,
  177

\bibitem[{{Spitler} {et~al.}(2016){Spitler}, {Scholz}, {Hessels}, {Bogdanov},
  {Brazier}, {Camilo}, {Chatterjee}, {Cordes}, {Crawford}, {Deneva}, {Ferdman},
  {Freire}, {Kaspi}, {Lazarus}, {Lynch}, {Madsen}, {McLaughlin}, {Patel},
  {Ransom}, {Seymour}, {Stairs}, {Stappers}, {van Leeuwen}, \&
  {Zhu}}]{Spitler+2016}
{Spitler}, L.~G., {Scholz}, P., {Hessels}, J.~W.~T., {et~al.} 2016, \nat, 531,
  202

\bibitem[{{Tendulkar} {et~al.}(2017){Tendulkar}, {Bassa}, {Cordes}, {Bower},
  {Law}, {Chatterjee}, {Adams}, {Bogdanov}, {Burke-Spolaor}, {Butler},
  {Demorest}, {Hessels}, {Kaspi}, {Lazio}, {Maddox}, {Marcote}, {McLaughlin},
  {Paragi}, {Ransom}, {Scholz}, {Seymour}, {Spitler}, {van Langevelde}, \&
  {Wharton}}]{Tendulkar+2017}
{Tendulkar}, S.~P., {Bassa}, C.~G., {Cordes}, J.~M., {et~al.} 2017, \apjl, 834,
  L7

\bibitem[{{Thornton} {et~al.}(2013){Thornton}, {Stappers}, {Bailes},
  {Barsdell}, {Bates}, {Bhat}, {Burgay}, {Burke-Spolaor}, {Champion}, {Coster},
  {D'Amico}, {Jameson}, {Johnston}, {Keith}, {Kramer}, {Levin}, {Milia}, {Ng},
  {Possenti}, \& {van Straten}}]{Thornton+2013}
{Thornton}, D., {Stappers}, B., {Bailes}, M., {et~al.} 2013, Science, 341, 53

\bibitem[{{Vedantham} {et~al.}(2016){Vedantham}, {Ravi}, {Hallinan}, \&
  {Shannon}}]{Vedantham+2016}
{Vedantham}, H.~K., {Ravi}, V., {Hallinan}, G., \& {Shannon}, R.~M. 2016, \apj,
  830, 75

\bibitem[{{Weiler} {et~al.}(1990){Weiler}, {Panagia}, \&
  {Sramek}}]{WeilerPS1990}
{Weiler}, K.~W., {Panagia}, N., \& {Sramek}, R.~A. 1990, \apj, 364, 611

\bibitem[{{Williams} \& {Berger}(2016)}]{WilliamsB2016}
{Williams}, P.~K.~G., \& {Berger}, E. 2016, \apjl, 821, L22

\bibitem[{{Yang} {et~al.}(2017){Yang}, {Luo}, {Li}, \& {Zhang}}]{Yang+2017}
{Yang}, Y.-P., {Luo}, R., {Li}, Z., \& {Zhang}, B. 2017, \apjl, 839, L25

\bibitem[{{Yang} \& {Zhang}(2017)}]{YangZ2017}
{Yang}, Y.-P., \& {Zhang}, B. 2017, \apj, 847, 22

\bibitem[{{Zanardo} {et~al.}(2014){Zanardo}, {Staveley-Smith}, {Indebetouw},
  {Chevalier}, {Matsuura}, {Gaensler}, {Barlow}, {Fransson}, {Manchester},
  {Baes}, {Kamenetzky}, {Laki{\'c}evi{\'c}}, {Lundqvist}, {Marcaide},
  {Mart{\'{\i}}-Vidal}, {Meixner}, {Ng}, {Park}, {Sonneborn}, {Spyromilio}, \&
  {van Loon}}]{Zanardo+2014}
{Zanardo}, G., {Staveley-Smith}, L., {Indebetouw}, R., {et~al.} 2014, \apj,
  796, 82

\end{thebibliography}

\clearpage

\end{document}